# Field dependent competing magnetic ordering in multiferroic $Ni_3V_2O_8$

A. K. Singh[1], D. Jain[2], V. Ganesan[2], A. K. Nigam[3] and S. Patnaik[1]

[1] School of Physical Sciences, Jawaharlal Nehru University, New Delhi 110067, India

[2] UGC-DAE Consortium for Scientific Research, University Campus, Khandwa Road, Indore 452017, India

[3] Tata Institute of Fundamental Research, Bombay 400 005, India

# Abstract

The geometrically frustrated magnet $Ni_3V_2O_8$ undergoes a series of competing magnetic ordering at low temperatures. Most importantly, one of the incommensurate phases has been reported to develop a ferroelectric correlation caused by spin frustration. Here we report an extensive thermodynamic, dielectric and magnetic study on clean polycrystalline samples of this novel multiferroic compound. Our low temperature specific heat data at high fields up to 14 Tesla clearly identify the development of a new magnetic field induced phase transition below 2 K that shows signatures of simultaneous electric ordering. We also report temperature and field dependent dielectric constant that enables us to quantitatively estimate the strength of magneto-electric coupling in this *improper* ferroelectric material.

## Introduction

Multiferroic materials exhibit juxtaposition of electric and magnetic ordered phases and thus give rise to the long cherished possibility of switching electric polarization by application of magnetic field and magnetization by electric field. This has formed the basis of many of the futuristic device propositions based on electric control of spin polarized current and voltage controlled magnetic storage media [1-4]. While the subject is old, the current excitement stems from two recent developments; the discovery of *improper* ferroelectricity in geometrically frustrated magnetic systems [5-7] and a better understanding on the microscopic origin of multiferrocity in $BiFeO_3$ and $YMnO_3$ [8-9]. In this paper we report an extensive thermodynamic, magnetic and dielectric study of one of the keenly studied multiferroic of recent times; the geometrically frustrated system of $Ni_3V_2O_8$ (NVO). The magnetic field – temperature (H-T) phase diagram of NVO has been studied from combined thermodynamic [10-11], magnetic [10-14], neutron scattering [10, 11, 15], optical [15], and dielectric [16-17] measurements. The consensus is that at low temperature NVO undergoes at least four distinct magnetic phase transitions below 12 K, above which a robust paramagnetic phase dominates. These phases are the i) high temperature incommensurate phase (HTI), ii) low temperature incommensurate phase (LTI), and iii) two canted antiferromagnetic phases (C and C'). There have been some reports indicating a possibility of a fifth magnetic field induced phase at temperature below 2.5 K from optical measurements [18, 19]. However these were open to interpretation as no thermodynamic features were confirmed. In this letter we present the synthesis and characterization of phase pure polycrystalline $Ni_3V_2O_8$ samples prepared by solid state reaction processing. The specific heat data below ~ 2 K in the presence of magnetic field unambiguously shows the onset of a new magnetic field induced phase that exhibits clear signatures of a robust electric correlation.

Structurally NVO belongs to the compounds with general chemical formula $M_3V_2O_8$ (M = $Ni^{2+}$, $Co^{2+}$, $Zn^{2+}$, and $Cu^{2+}$) [12-14, 20-22]. Some of these materials are characterised by triangular lattices and short range antiferromagnetic interactions as is the case with NVO [23, 24]. Here the magnetic lattice is based on staircase Kagome net where magnetic $Ni^{2+}$ (S=1, $d^8$) is arranged anisotropically (Fig. 1a). The planes that contain the edge sharing $NiO_6$ octahedra are not flat (as in regular Kagome lattice) but buckled forming a staircase-like structure. These magnetic layers are separated by non-magnetic $VO_4$ tetrahedra. As a consequence the geometric frustration is reduced allowing long range magnetic order. As elucidated in Fig 1a, there are two crystallographically inequivalent magnetic sites $Ni_c$ (cross tie) and $Ni_s$ (spine) in the Kagome plane. The isosceles triangles are formed by $Ni_s$-$Ni_c$-$Ni_s$ and the ratio of the $Ni_s$-$Ni_s$ and $Ni_c$-$Ni_s$ bond length is estimated at ~1.01. The positions of the two kinds of $Ni^{2+}$ which has been referred to as “spine” and “cross tie” are shown with blue and light green spheres. The deviation from the ideal Kagomé geometry introduces several new interactions that relieve the frustration of underlying Kagomé antiferromagnets in unexpected ways [10]. These smaller interactions play a crucial role to trace an explanation of the microscopic origins of multiferrocity observed in NVO.

**Experiments**

Polycrystalline samples of $Ni_3V_2O_8$ were prepared by solid-state reaction at ambient pressure. High purity nickel oxide NiO and vanadium oxide $V_2O_5$ were used as starting materials. The reagents were mixed thoroughly in stoichiometric proportion, compacted and calcinated at 800 °C for 16 hours in dense $Al_2O_3$ crucibles. The reacted product was ground, compacted, and reheated at 850 °C for 16 hours. The product was characterized using X-ray powder diffraction (XRD) by using BRUKER D-8 advanced diffractometer with Cu K α (λ = 1.5418 Å) radiation and with a scan step width of 0.02°. Composition analysis was performed using an electron diffraction energy dispersive x-ray spectroscopy (EDAX). The capacitance (C) and the loss factor (tanδ) were measured with a QUADTECH 1920 precision LCR meter. The dielectric measurements were performed in a 'Cryogenic' 8 T cryogen-free magnet system with an attached variable temperature insert [25, 26]. The dc magnetization was measured using Quantum Design SQUID. Heat capacity measurements were performed on sintered powder sample using a relaxation technique for temperatures below 15K and fields upto 14 Tesla.

## Results and discussion

Figure 1b shows the observed XRD pattern where all the major peaks of $Ni_3V_2O_8$ are identified and indexed. The sample exhibits single phase orthorhombic structure with space group *Cmca*. The lattice parameters for $Ni_3V_2O_8$ are calculated to be a = 5.9178 Å, b = 11.3652 Å and c = 8.2896 Å. EDAX measurement (Fig. 1c) confirms that there is no other metal element except for Ni and V. Within the experimental error the molar ratio of Ni: V: O is approximated to be 3:2:8, which is in general agreement with chemical formula. The dielectric constant as a function of temperature (at 5 kHz) is shown in Fig. 2. A peak in dielectric constant at $T_{HL}$ = 6.9 K indicates the onset of ferroelectric order corresponding to the onset of low temperature incommensurate phase. In the inset of Fig. 2 dielectric constant in zero field and in the presence of 3 Tesla magnetic field (at 5 kHz) are shown. In the presence of magnetic field, the peak in the dielectric constant becomes sharper as compared to the case in zero field. This is a clear signature of magnetoelectric coupling in this geometrically frustrated phase. The calculated magnetocapacitance ((ε(H, T) - ε(H=0, T)/ ε(H=0, T)) ×100) is about 0.3 which is small as compared to other frustrated magnetic systems such as orthorhombic $HoMnO_3$ and $YMnO_3$ [27, 28]. We note that the degree of frustration which is quantified as $\theta/T_H$ (θ being the Curie-Weiss intercept and $T_H$, the magnetic ordering temperature) for $Ni_3V_2O_8$ is ~3.1 which is large as compared to orthorhombic $YMnO_3$ (~1.27) and $HoMnO_3$ (~0.5). In case of $YMnO_3$ the dielectric constant is decreases by 2% and in case of $HoMnO_3$ this decrease is 8 % in the presence of 7 T magnetic field [27]. While a microscopic understanding on this would need detailed study of the magnetic structure of each of these materials, we can qualitatively observe that the degree of frustration is inversely proportional to the magneto-dielectric effect. Most importantly, we find that there is clear evidence of the onset of another electric ordering below ~ 2 K. Hitherto it was suggested [13] that only the LTI phase is ferroelectric so the origin of the possible second ferroelectric transition in the low temperature antiferromagnetic phase (C') needs to be assessed. We emphasize that we have done the zero field dielectric measurement at different frequencies and this lower temperature increase in dielectric constant is consistent and reproducible. At lower frequencies both the anomalies at $T_{HL}$= 6.4 K and T ~ 2 K in the dielectric constant is higher than that measured at higher frequencies. This onset of re-entrant long-range electric correlation persists up to 1.6 K which is the lowest temperature in our measurement.

The magnetization data for $Ni_3V_2O_8$ in applied field of 0.1 T are shown in fig.3 a. The inverse susceptibility $1/\chi$ data are shown in the inset of fig 3a. The high temperature inverse susceptibility data show excellent linearity and as shown in inset of fig 3a. departure from linearity could be observed around ~ 10 K with extrapolated Curie –Weiss intercept $\theta$ ~ -11 K. The negative value of $\theta$ indicates dominant role of antiferromagnetic interaction between $Ni^{2+}$ spins amongst all other competing interactions. From magnetization measurement on single crystal NVO, a first order magnetic transition is observed across the low temperature incommensurate to canted antiferromagnetic phase boundary (~ 4 K), where strong anisotropic features set in [17]. It is understood that the antiferromagnetic alignment occurs along **a** axis with canting towards **c** axis. This canting leads to sharp increase in magnetization along **c**-axis. At ~ 2.8 K the magnetic state crosses over to the second canted antiferromagnetic phase with different anisotropies. This is indicated as a downturn in overall magnetization in our polycrystalline data. The canting along c-axis also gives rise to substantial hysteresis at 2 K as observed in fig 3b. At 5 K, above the CAF phase, no hysteresis is observed in the incommensurate phase.

The low temperature heat capacity data for $Ni_3V_2O_8$ is shown in different panels of Fig. 4. The zero field heat capacity as a function of temperature is shown in the first panel. It shows four anomalies at 2.8, 4.0, 6.4, and 9.3 K corresponding to four magnetic phase transitions. These four peaks in NVO specific heat correspond to the development of complex spin structures as established by neutron diffraction measurements [11, 15]. The peaks at T = 9.3 K ($T_{PH}$) and 6.4 K ($T_{HL}$) correspond to second order phase transitions. Both these phases are magnetically incommensurate states. There is another distinct peak at T= 4.0 K ($T_{LC}$) corresponding to a first order magnetic phase transition consequent upon a progression from an incommensurate to canted antiferromagnetic phase. In LTI phase (between 6.4 and 4 K), the spine and cross tie spin rotate within a-b planes leading to remarkable ferroelectric behaviour along with field dependent electric polarization [13, 17]. This is well supported by our magneto-dielectric measurement. Below $T_{LC}$ there are two commensurate phases C and C′. The temperature corresponding to these commensurate phases are $T_{LC}$ = 4 K and $T_{CC'}$ = 2.8 K. The magnetic structure of this fourth transition at 2.8 K ($T_{CC'}$) is yet to be fully understood. We also note that as compared to separation of $T_{PH}$ - $T_{HL}$ and $T_{HL}$-$T_{LC}$, the separation of $T_{LC}$-$T_{CC'}$ is small and this is related to the fact that there exist two subsystems of $Ni^{2+}$ ions. One $Ni^{2+}$ corresponds to cross-tie site ($Ni_i$) and second corresponds to spine sites ($Ni_s$) and the number of spine sites is twice that of cross sites. In the temperature region, $T_{CC'}$ < T < $T_{LC}$ the subsystem of $Ni_i^{2+}$ primarily constitute the incommensurate low temperature phase and the $Ni_s^{2+}$ ions begin to transit to this phase at $T_{CC'}$. With the application of magnetic field these two transitions are expected to merge. This hypothesis appears to be true in our field dependent specific heat data shown in panels for H = 0, 9 and 14 Tesla. Applying 1 T magnetic field produces substantial broadening in all the peaks ($T_{CC}'$, $T_{LC}$, $T_{HL}$, $T_{PH}$), but does not significantly shift the transition temperatures. With the application of higher magnetic fields of 9 T and 14 T the two peaks at the lower temperature region $T_{CC}'$, and $T_{LC}$ merge together. The other two peaks $T_{HL}$, and $T_{PH}$ are distinct even at 9 T magnetic field but the broadness of these peaks increases. At 14 T only one peak remains. Next we calculate the entropy of $Ni_3V_2O_8$ system by integrating C/T versus T plots up to 13 K. For 0 T, the entropy calculated for $Ni_3V_2O_8$ system is 6.10 J/mol-K. This value is approximately equal to the 67% of value expected for spin-1 system. As we apply magnetic field, the entropy of the system decreases quite consistently. The value of calculated entropy after the application of 1 T, 9 T, and 14 T magnetic fields in the temperature range 0 to 13 K are 6.03, 5.47, and 5.10 J/mol-K. Further we observe the emergence of a low temperature field induced phase for H= 1T and 9T below 2 K that disappears both at 0T and 14 T. Because of the temperature limitation of our measurement system, we could not access further low temperature and this low temperature peak is not fully developed but its onset is effectively established. This has also been indicated in our dielectric data shown in Fig. 2. This result gives thermodynamic credence to what was reported from theoretical calculations and optical measurements [19].

In conclusion high quality polycrystalline sample of the Kagomé staircase lattice $Ni_3V_2O_8$ have been prepared. The magnetic measurements reveal short range ferromagnetic correlations below $T_{CC}'$ corresponding to canted antiferromagnetic phases. Field dependent heat capacity clearly establishes the merging of two peaks corresponding to two canted antiferromagnetic phases below 4 K. Most importantly we observe the onset of a new low temperature (< 2 K) field induced phase that is accompanied by simultaneous development of a robust electric ordering. In essence we provide evidence of a re-entrant multiferroic phase in this magnetically rich frustrated spin system.

**Acknowledgement**

We thank the Department of Science of Technology, Government of India, for the financial support under the FIST program to JNU, New Delhi. A.K.S. and DJ would like to thanks CSIR, India for the fellowship. We acknowledge DST for SQUID, at Department of Physics, Indian Institute of Technology, Delhi, India and 14T PPMS system at UGC-DAE-CSR, Indore. We acknowledge very useful discussions with A. K. Rastogi (JNU) on the magnetization data.

**Figure Caption**

Fig.1(a). Crystal structure of $Ni_3V_2O_8$ showing spin-1 $Ni^{2+}$ spine sites in blue and cross tie sites in light green. Competing interaction between spine and cross tie $Ni^{+2}$ and anisotropies yield the complex magnetic structure in NVO. (b) Room temperature X- ray diffraction patterns of $Ni_3V_2O_8$ that establish orthorhombic crystal structure with space group *Cmca*. All the peaks are identified and indexed. (c) Energy dispersive x-ray spectroscopy exhibits the elements Ni, V and O in the specimen. Star (*) marked peaks are from the carbon tape used to glue the sample.

Fig.2. Temperature dependence of dielectric constant measured at 5 kHz (■) is shown. The large peak in dielectric constant at $T_L$ = 6.4 K shows the onset of ferroelectric order in NVO. Below 2.1 K there is also a large anomaly in ε suggesting possibility of another ferroelectric phase. Inset shows zero field (■) and 3 T (●) magnetic field at f = 5 kHz.

Fig.3(a). Temperature dependent magnetization of $Ni_3V_2O_8$ at applied field of 1000 Oe. Inset shows the corresponding inverse susceptibility plot. (b). Magnetization is plotted as a function of magnetic field at T = 2 K. Inset shows M-H loop at 5 K that does not show hysteresis.

Fig.4. Low temperature specific heat of $Ni_3V_2O_8$ at H = 0 T, 1 T, 9 T, and 14 T magnetic fields. In zero field measurement four sharp peaks are observed at 2.6, 4.0, 6.4, and 9.3 K. At H = 1T and H = 9T, emergence of a field induced phase is observed (marked by arrow).

**Fig. 1 a**

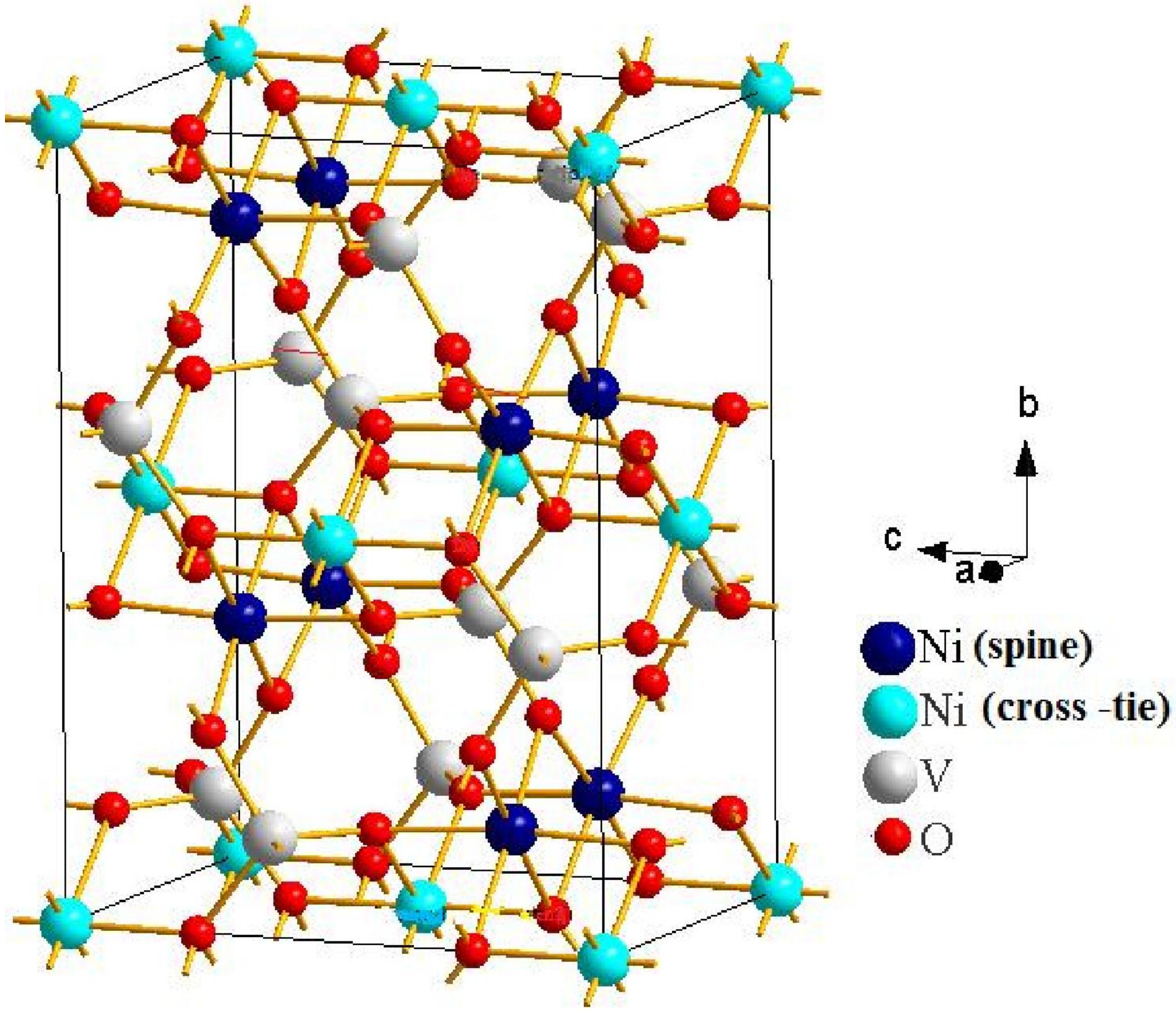

Fig. 1b

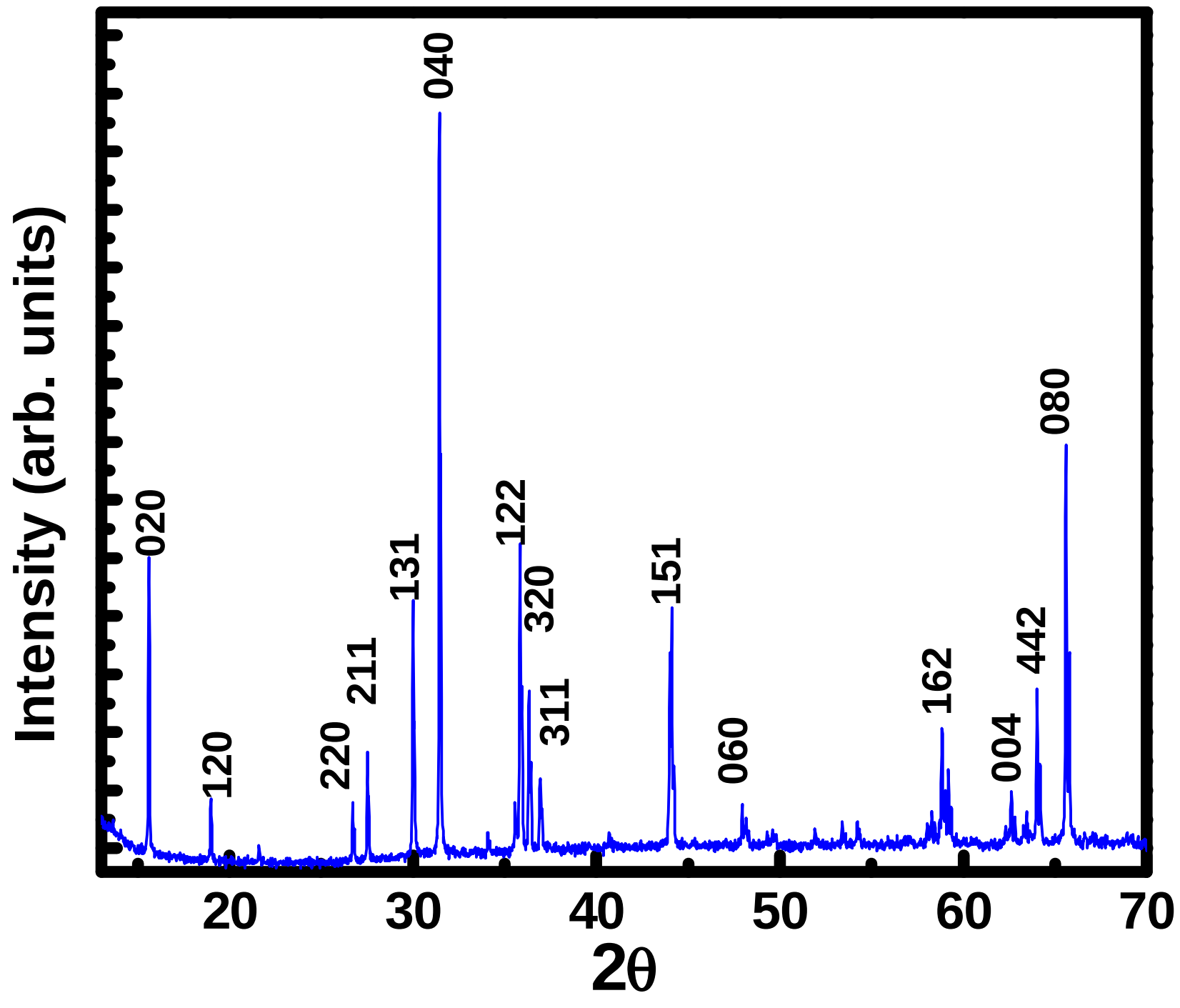

Fig 1c

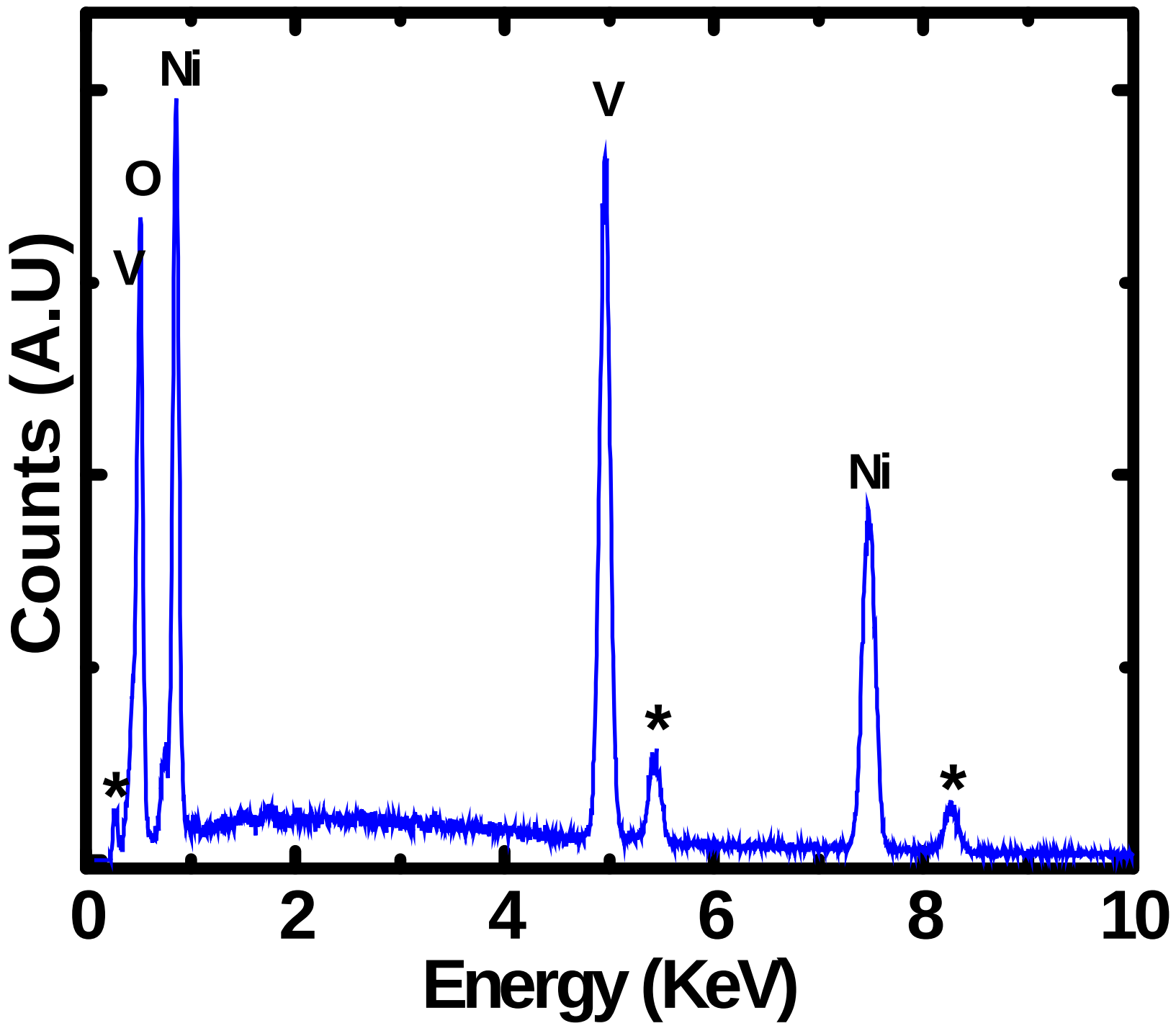

**Fig.2**

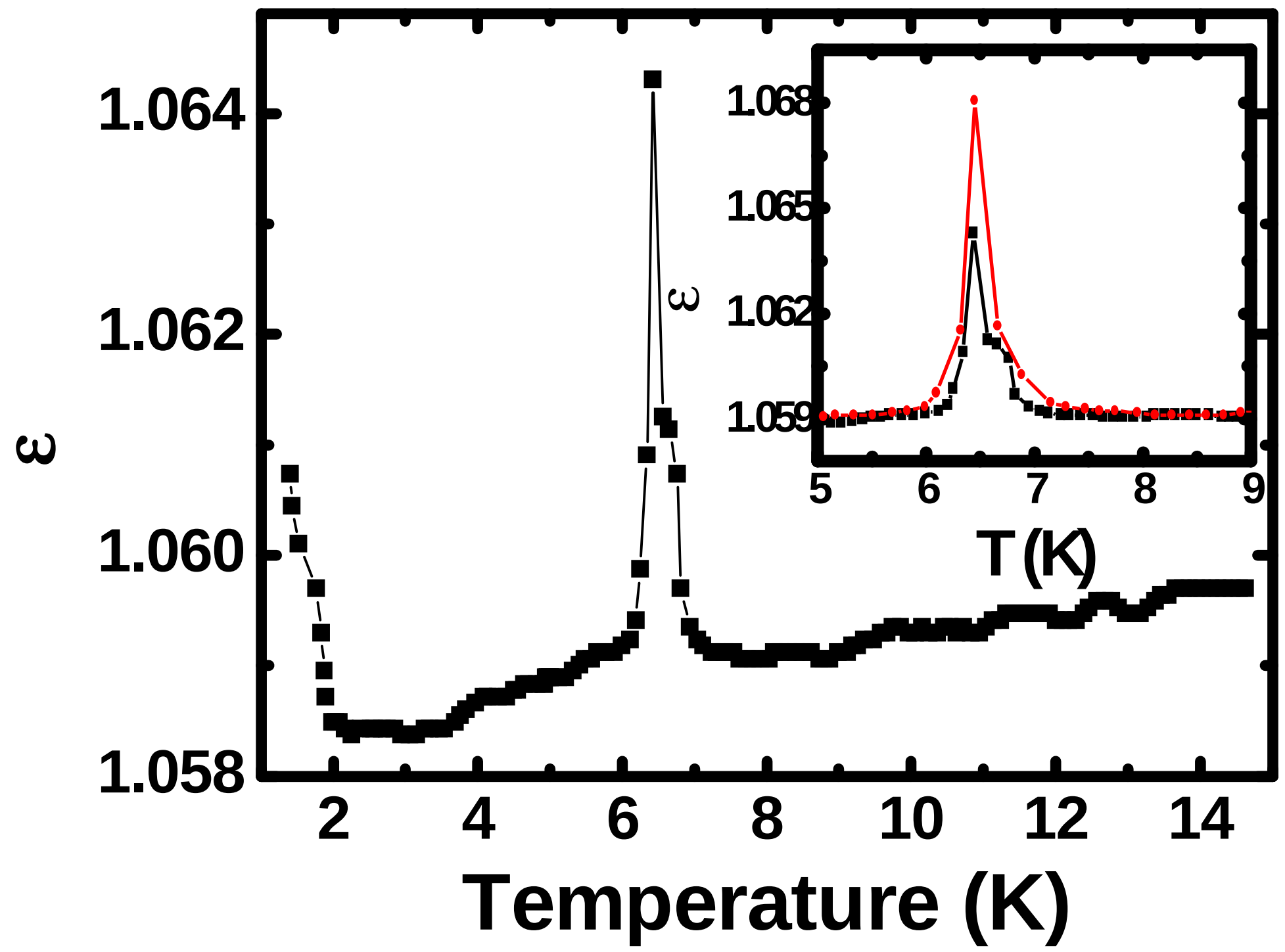

**Fig.3 (a)**

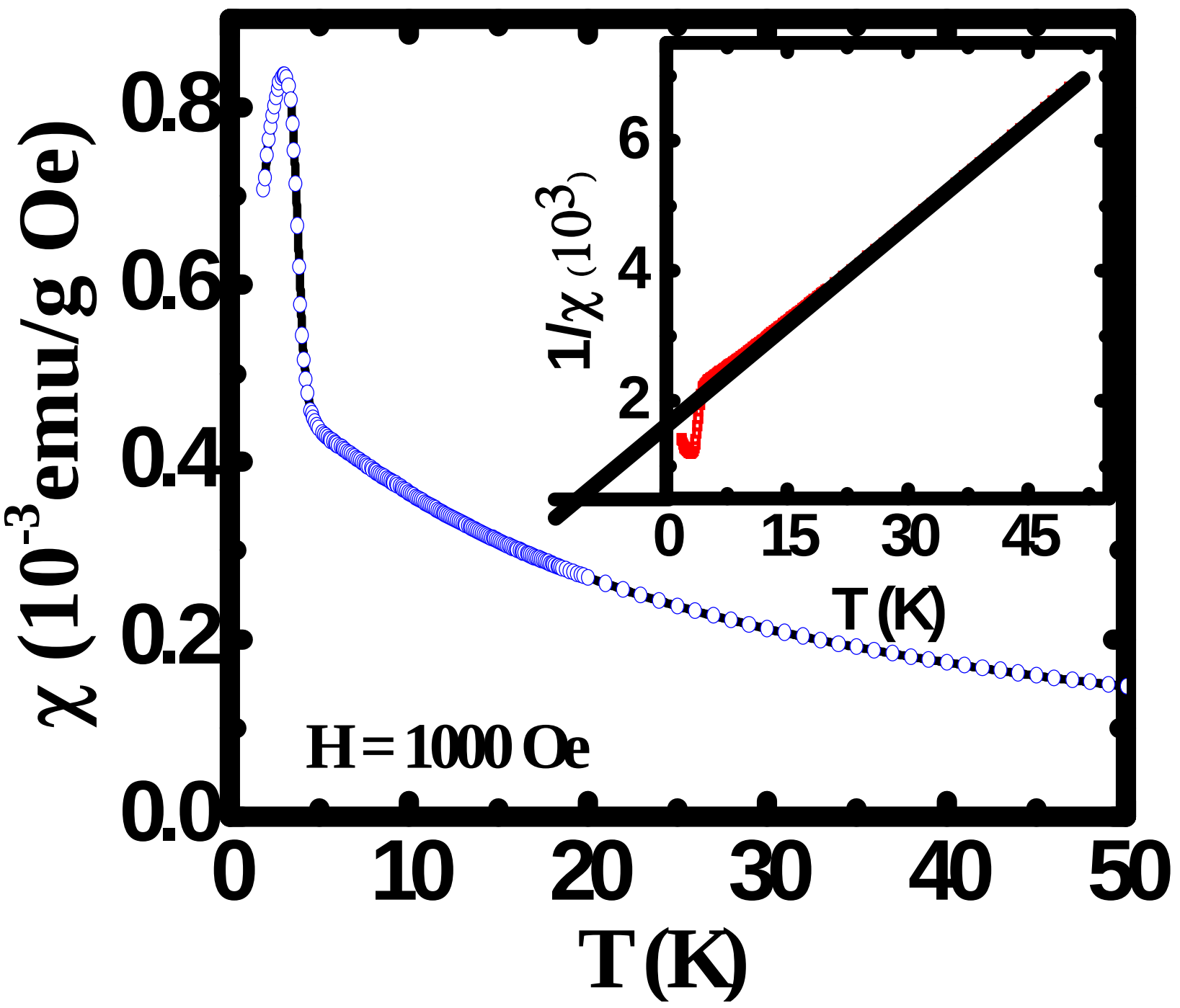

**Fig.3 (b)**

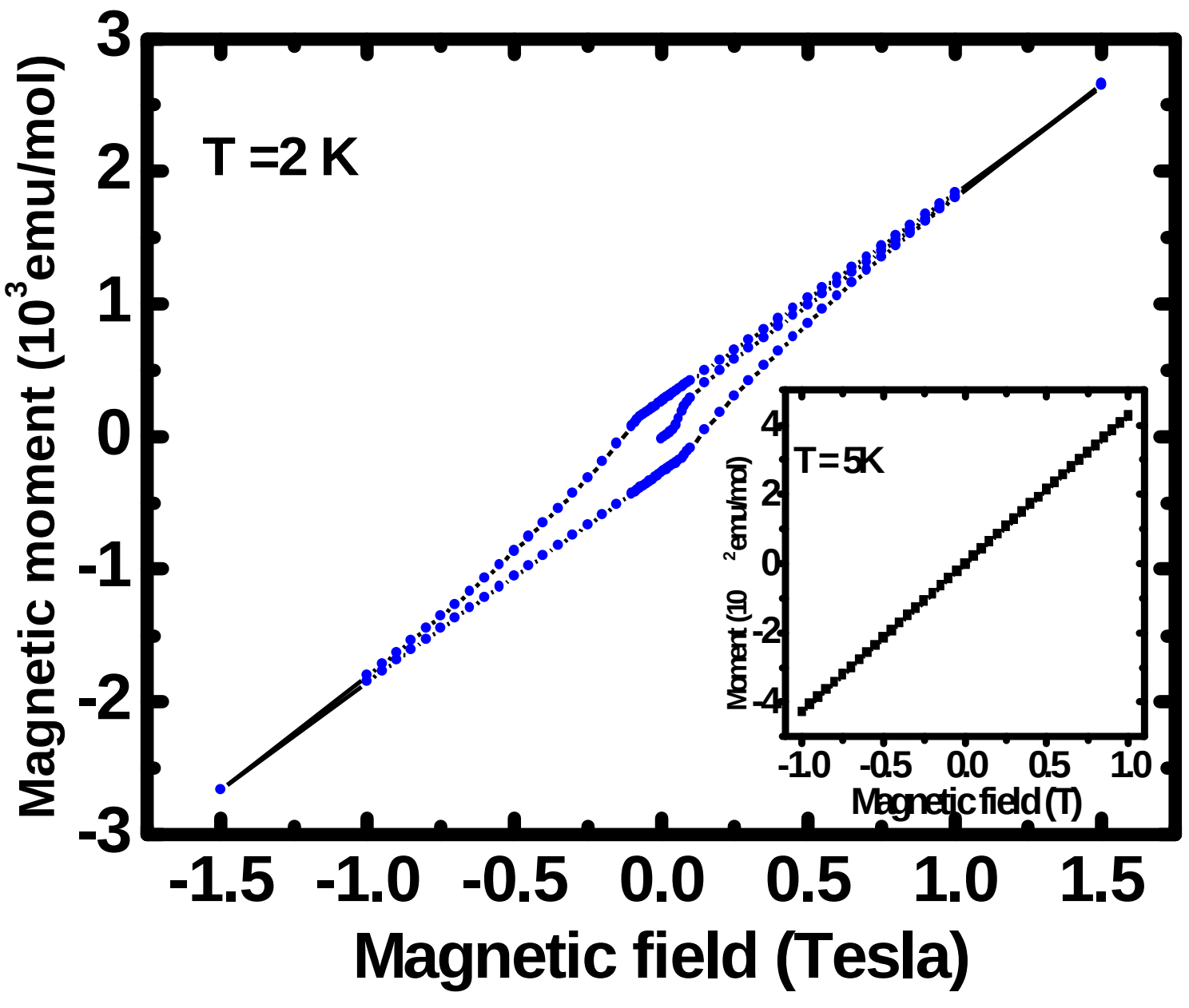

Fig. 4

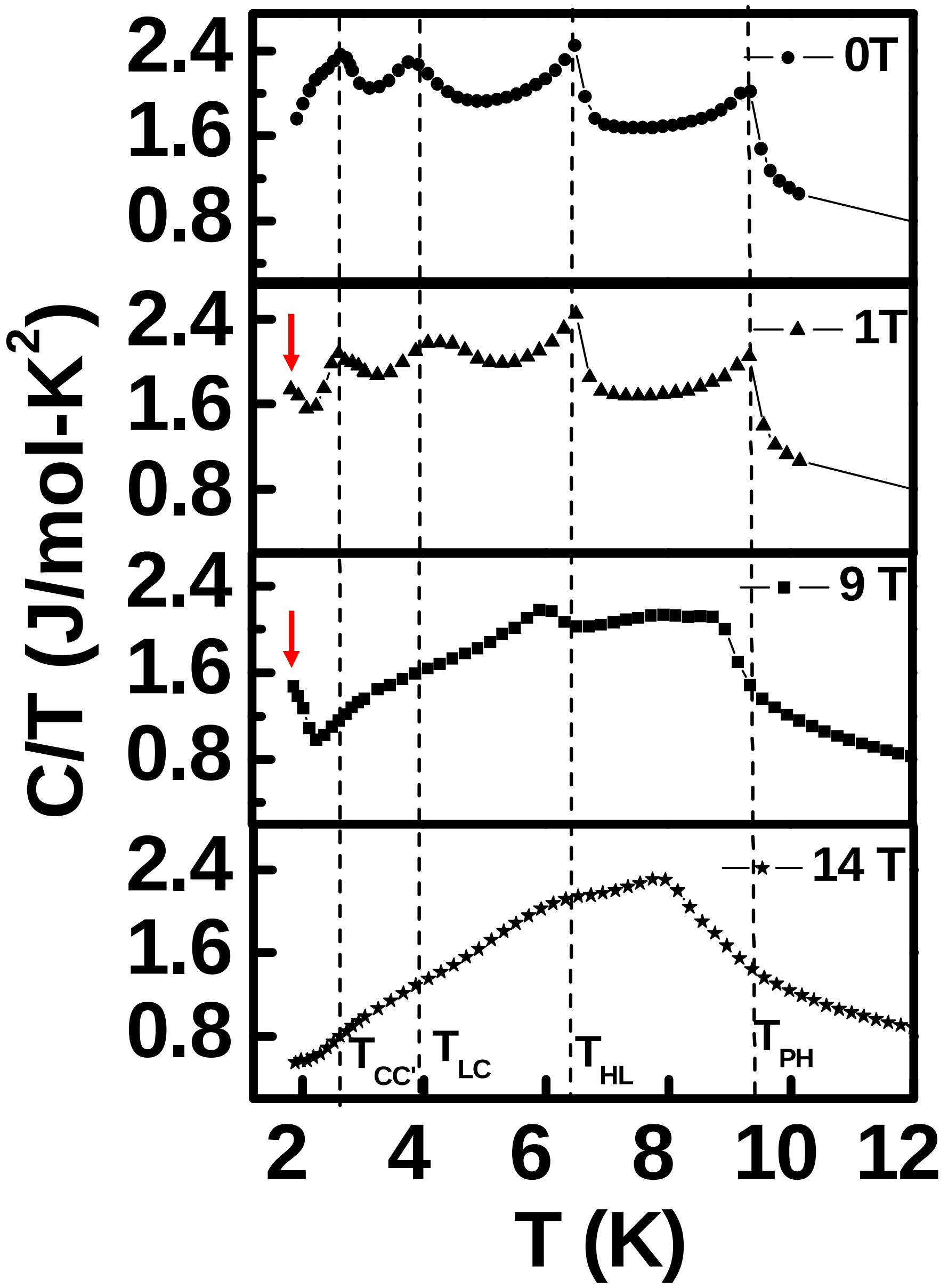